\documentstyle[preprint,epsf,aps]{revtex}
\begin{document}
\title{Spin-Peierls transition in an anisotropic two-dimensional XY model}
\author{Qingshan Yuan$^{1,2}$, Yumei Zhang$^2$ and Hong Chen$^2$}
\address{$^1$Max-Planck-Institut f\"{u}r Physik komplexer Systeme, 
N\"{o}thnitzer Str.38, 01187 Dresden, Germany\\
$^2$Pohl Institute of Solid State Physics, Tongji University, 
Shanghai 200092, P.R.China}
\maketitle
\begin{abstract}
The two-dimensional Jordan-Wigner transformation is used to investigate the 
zero temperature spin-Peierls transition for an anisotropic two-dimensional 
XY model in adiabatic limit. The phase diagram between the dimerized (D) 
state and uniform (U) state is shown in the parameter space of dimensionless 
interchain coupling $h$ $(=J_{\perp}/J)$ and spin-lattice coupling 
$\eta$. It is found that the spin-lattice coupling $\eta$ must exceed some 
critical value $\eta_c$ in order to reach the D phase for any finite $h$. 
The dependence of $\eta_c$ on $h$ is given by $-1/\ln h$ for $h\rightarrow 0$ 
and the transition between U and D phase is of first-order for at least 
$h>10^{-3}$. 
\end{abstract}


The fact that the $S=1/2$ Heisenberg spin chain is unstable towards 
dimerization when coupled to an elastic lattice is known as spin-Peierls (SP) 
transition \cite{Cross}. This occurs because dimerization opens a gap in 
the spin excitation spectrum and lowers the total magnetic energy by a greater
amount than the increase in elastic energy due to lattice deformation. Such a
transition was first suggested by analogy with the conventional Peierls
transition in linear conducting chains and later observed in several organic
compounds such as MEM-(TCNQ)$_2$ \cite{Huizinga}. It has attracted renewed
attention since the discovery of inorganic compound CuGeO$_3$ in 1993
\cite{Hase}.

In the adiabatic limit, it is well understood for exactly one-dimensional (1D)
XY or Heisenberg model that the dimerized (D) state, relative to the uniform 
(U) one, is always stable for arbitrarily weak spin-lattice coupling
in the ground state \cite{Cross}. In presence of some extra mechanisms these 
two states may compete leading to interesting phase diagram.
Up to now there are several mechanisms which were found to destroy the D 
phase, for example, small Ising anisotropy \cite{Inagaki}. Another mechanism is
quantum lattice fluctuation which was intensely studied 
recently \cite{Caron,Bursill}. 
Beyond adiabatic approximation it was found 
that the spin-phonon coupling must be larger than some nonzero critical value 
for the SP instability to occur \cite{Caron,Bursill}.
The interchain exchange coupling is also a possible mechanism 
\cite{Fukuyama,Dobry,Azzouz2}. 
This mechanism is often not negligible for real SP materials, especially
for the material CuGeO$_3$ where a relatively large interchain coupling, 
about $J_{\perp}/J=0.1$ ($J$ represents the intrachain exchange) was found
experimentally \cite{Nishi}. Theoretically the importance of
the interchain coupling was also stressed by many authors
in description of CuGeO$_3$ \cite{Bouzerar,Rosner,Knetter}. 
Therefore it is necessary to 
study the SP transition for a general {\it quasi}-1D system with inclusion of
interchain coupling. This should be discussed by answering a central
problem how the interchain coupling modifies the dimerized ground state 
of the exactly 1D case. 

In this work we propose to study the above problem through a two-dimensional 
(2D) XY model with anisotropic intrachain and interchain coupling, and leave
the generalization to the case of the corresponding Heisenberg model for later
study. The reason is two-fold: i. the problem is simplified; especially 
the 1D limit can be exactly solved. ii. the XY model is believed to
contain the essential 
elements for the spin-Peierls transition \cite{Caron}. We point out that 
the similar problem based on the Heisenberg model was studied previously 
by some authors \cite{Fukuyama,Dobry}. However, they all treated 
the interchain coupling within the so called chain mean-field theory and 
actually still studied a 1D system. Usually the effect of interchain coupling 
is underestimated in this way. Here, the two-dimensional (2D) Jordan-Wigner 
(JW) transformation \cite{Wang1,Wang2,Azzouz1,Ji,Azzouz2} is adopted so that 
the intrachain and interchain exchanges may be treated on an equal footing 
from the beginning.

In adiabatic approximation, we begin with the following anisotropic 2D 
XY model on a square lattice:
\begin{eqnarray}
H & = & J\sum_{i,j} (1+\alpha u_{i,j})(S_{i,j}^x\cdot S_{i+1,j}^x+
S_{i,j}^y\cdot S_{i+1,j}^y)
+J_{\perp}\sum_{i,j} (S_{i,j}^x\cdot S_{i,j+1}^x+S_{i,j}^y\cdot S_{i,j+1}^y)
+{K\over 2}\sum_{i,j}u_{i,j}^2\ ,
\end{eqnarray}
where $S_{i,j}^{x/y}$ represents the $x/y$ component of the spin-$1/2$ 
operator at site $\vec{r}_{i,j}=i\vec{x}+j\vec{y}$ ($\vec{x}$, $\vec{y}$ are 
unit vectors along the $x$-
and $y$-axis, respectively), $J(J_{\perp})$ is the intra(inter)chain exchange 
coupling along the $x(y)$-axis, $\alpha$ represents spin-lattice coupling 
constant and $K$ is the intrachain elastic constant. Here the lattice 
displacement is assumed as $u_{i,j}=(-1)^{i+j}u$, i.e., corresponding
to a wave vector $(\pi,\pi)$, which was found 
to be the case in CuGeO$_3$ by neutron-diffraction measurements \cite{Hirota}.
In the following we take $J$ as unit of energy and use several dimensionless 
parameters: the dimerization parameter $\delta=\alpha u$, the interchain 
coupling $h=J_{\perp}/J$ and the spin-lattice coupling $\eta=\alpha^2 J/K$.
The total number of lattice sites is $N$.

The spin operators may be represented as fermions by use of the 2D JW 
transformation, which has the advantage that all spin commutation relations, 
as well as the
so called spin on-site exclusion principle are automatically preserved
\cite{Wang2}. This method has been proved well 
for application to real materials \cite{Wang2,Azzouz2}. For concreteness, the
following formulas are adopted \cite{Azzouz1,Ji}
\begin{eqnarray}
S_{i,j}^{-} =  c_{i,j}\exp \{i\phi_{i,j}\},\ 
S_{i,j}^{+} =  \exp \{-i\phi_{i,j}\}c_{i,j}^{\dagger},\ 
S_{i,j}^{z} =  c_{i,j}^{\dagger}c_{i,j}-1/2 \label{JW}
\end{eqnarray}
with
\begin{eqnarray}
\phi_{i,j} & = & \pi\left[ \sum_{l=0}^{i-1}\sum_{m=0}^{\infty}
n_{l,m}+\sum_{m=0}^{j-1}n_{i,m}\right]\ ,\nonumber
\end{eqnarray}
where $c_{i,j}^{\dagger} (c_{i,j})$ is creation (annihilation) operator 
for a spinless electron at $\vec{r}_{i,j}$ and $n_{i,j}=c_{i,j}^{\dagger}
c_{i,j}$ is the particle number operator. The original Hamiltonian can 
then be transformed into:
\begin{eqnarray}
H & = & \sum_{i,j}[1+(-1)^{i+j}\delta](e^{i\psi_{i,j}}
c_{i,j}^{\dagger}c_{i+1,j}+{\rm h.c.})/2
+h\sum_{i,j}(c_{i,j}^{\dagger}c_{i,j+1}+{\rm h.c.})/2
+N\delta^2/2\eta \ ,\label{H_JW}
\end{eqnarray} 
where the phase factor $e^{i\psi_{i,j}}=e^{i(\phi_{i+1,j}-\phi_{i,j})}$ 
describes an effective gauge field acting on spinless fermions.
Note that the Hamiltonian (\ref{H_JW}) is exact and it may recover to the
corresponding one derived from 1D JW transformation as $h\rightarrow 0$.
When going on one has to treat the phase factor approximately. Following 
previous work, we select a
configuration (see Figs. 1 in Refs. \cite{Wang1,Wang2,Ji}) where the phase 
factor $e^{i\psi_{i,j}}$ has alternative values $e^{i\pi}$ and $1$, i.e., it 
varies with $(-1)^{i+j}$ (note that this coincides with the above dimerization 
pattern). This configuration ensures that each elementary plaquette encloses 
a net flux of $\pi$ \cite{Wang1,Wang2}. Then the Hamiltonian can be rewritten 
as follows in terms of fermion operators $e$ and $f$ corresponding
to the two sublattices A and B respectively 
(constants irrelevant to $\delta$ are ignored)
\begin{eqnarray}
H & = & {1\over 2}\sum_{\vec{r}_{i,j}\in A}[ -(1-\delta)
(f_{i-1,j}^{\dagger}e_{i,j}+{\rm h.c.})
+(1+\delta)(e_{i,j}^{\dagger}f_{i+1,j}+{\rm h.c.})\nonumber\\
& &\ \ \ \ \ \ \ \ +h(f_{i,j-1}^{\dagger}e_{i,j}+e_{i,j}^{\dagger}f_{i,j+1}
+{\rm h.c.})]+N\delta^2/2\eta \ . \label{H}
\end{eqnarray}
In the following we want to study the SP transition at $T=0$ based on the 
above Hamiltonian. Depending on the value $\delta^*$ which minimizes the 
ground state energy, the system may be in
D phase ($\delta^*\neq 0$) or U phase ($\delta^* =0$) in the 
parameter space of interchain coupling $h$ and spin-lattice coupling $\eta$. 

\begin{figure}[htb]
\epsfxsize=8cm
\epsfysize=6cm
\centerline{\epsffile{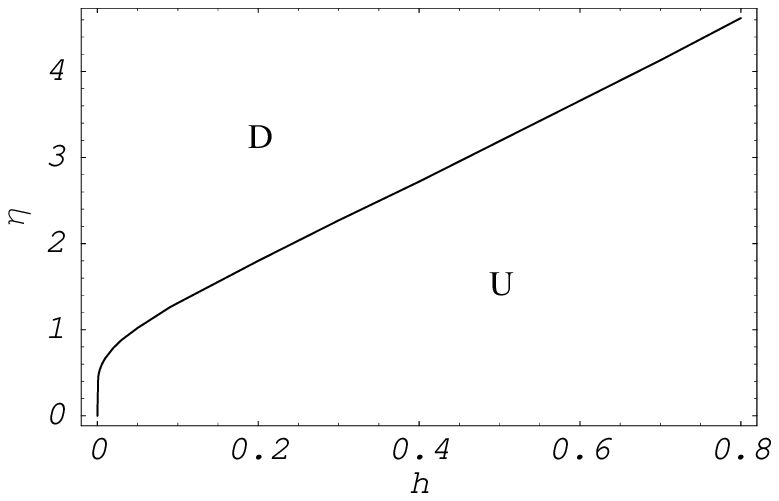}}
\caption{The D/U phase diagram for the anisotropic 2D XY model in the 
parameter space of 
interchain coupling $h$ and spin-lattice coupling $\eta$.}
\label{fig_PhDU}
\end{figure}

\begin{figure}[htb]
\epsfxsize=8cm
\epsfysize=6cm
\centerline{\epsffile{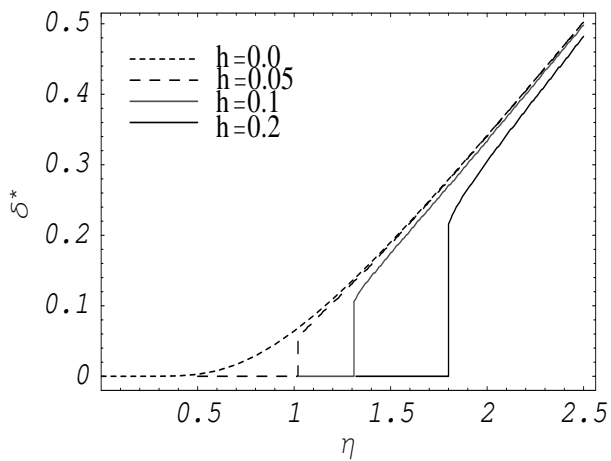}}
\caption{The order parameter $\delta^*$ as a function of $\eta$ for $h=0,0.05,
0.1$ and $0.2$.}
\label{fig_OP}
\end{figure}

The Hamiltonian (\ref{H}) may be exactly diagonalized. The electronic 
spectrum is written as
$\varepsilon_{\bf{k}}=\pm \sqrt{(\delta \cos k_x+h \cos k_y)^2+
\sin ^2 k_x}$ within the (reduced) Brillouin zone: $-\pi <k_x\pm k_y\le \pi$, 
and the ground state energy is simply given by $E_{\rm GS}=-\sum_{\bf{k}}|
\varepsilon_{\bf{k}}|+N\delta^2/2\eta$. Before continuing we add a comment
here on the above used JW transformation. In the definition of $\phi_{i,j}$ 
in Eq. (\ref{JW}) the summation is selected over the lattice sites along the 
direction perpendicular to the 
chains so that the phase factor appears in the intrachain hopping term as 
shown in Eq. (\ref{H_JW}). One may also think to do the summation along the 
chain direction, i.e., re-define $\phi_{i,j}=\pi\left[ \sum_{m=0}^{j-1}
\sum_{l=0}^{\infty}n_{l,m}+\sum_{l=0}^{i-1}n_{l,j}\right]$, and then the phase factor will appear in the interchain (rather than intrachain) hopping term. 
With the similar treatment for this phase factor as done above one may obtain 
an alternative spectrum $\pm \sqrt{(\delta \sin k_x+h \sin k_y)^2+
\cos ^2 k_x}$. 
However, the ground state energy $E_{\rm GS}$ (i.e., the summation over 
the Brillouin zone) is not changed by the new spectrum, 
nor all the subsequent results derived from it. Similar discussion
was given in Ref. \cite{Azzouz1}. Then we go on to search for the value 
$\delta^*$ which may be obtained from the condition 
$\partial E_{\rm GS}/\partial \delta =0 $, yielding the following 
equation:
\begin{eqnarray}
\delta & = & {\eta \over 2\pi^2}\int _{0}^{\pi}{\rm d}k_x
\int _{k_x-\pi}^{\pi-k_x}{\rm d}k_y \frac{(\delta \cos k_x+h \cos k_y)
\cos k_x}{\sqrt{(\delta \cos k_x+h \cos k_y)^2+\sin ^2 k_x}}\ . \label{delta} 
\end{eqnarray}
It is interesting to notice that Eq. (\ref{delta}) will reduce to the
corresponding rigorous equation for the exact 1D case \cite{Beni} if we 
set $h=0$, which means, 
\begin{eqnarray}
1 & = & {\eta \over\pi}\int _{0}^{\pi\over 2}{\rm d}k_x {\sin ^2 k_x\over
\sqrt{\delta ^2\sin ^2 k_x+\cos ^2 k_x}}\ . \label{delta_h0} 
\end{eqnarray}
The Eq. (\ref{delta_h0}) always has a single 
nonzero solution for $\delta$ as long as $\eta >0$ \cite{Remark}, 
which predicts a dimerized ground state. For $h\neq 0$, Eq. (\ref{delta}) may 
give one or two nonzero solutions $\delta_i\ (i=1,2)$ except for the trivial 
solution $\delta=0$.
We need to consider the sign of $\partial ^2 E_{\rm GS}/\partial \delta ^2
|_{\delta=\delta_{i}}$, as well as compare the energies between 
at $\delta=\delta_i$ and at end points $\delta=0,1$ to obtain the actual 
$\delta^*$. Without details, we give the results shown in Figs. \ref{fig_PhDU} 
and \ref{fig_OP}. The Fig. \ref{fig_PhDU} gives the D/U phase diagram in the 
parameter space $(h,\eta)$. As expected, the interchain coupling tends to 
suppress the D phase. As long as the interchain coupling $h>0$, the 
spin-lattice coupling $\eta$ must exceed some 
critical value $\eta_c$ to reach the D phase and the value $\eta_c$ increases 
monotonically with $h$. Careful analysis shows that 
the critical value $\eta_c$ has the functional form $-1/\ln h$ in the region 
$h\rightarrow 0$, which increases starting from zero 
much faster than any power law. Moreover, we have found that the transition
from U to D phase with increasing $\eta$ is of first-order for at least 
$h>10^{-3}$ \cite{Supp}. As shown in Fig. \ref{fig_OP} we may see how the 
order parameter $\delta^*$ jumps from zero to finite value with increase of 
$\eta$ for several $h$ values. Although the transition is of second-order 
at the point $h=0$ as can be proven from Eq. (\ref{delta_h0}), 
we speculate that it becomes of first-order as long as $h>0$.

It is also interesting to notice that the excitation spectrum may be 
gapless even for finite dimerization in presence of the interchain 
coupling. From $\varepsilon_{\bf{k}}$, it is easy to find that a gap
is not fully opened at the Fermi surface until the dimerization $\delta$ 
reaches the same value as the interchain coupling $h$. Qualitatively this
gives the physical reason for the conditional SP transition when $h>0$ as shown
by Fig. \ref{fig_PhDU}: The incomplete opening of the gap by the dimerization
is unfavorable to an adequate reduction for the electronic energy
so that the gain from it is not always 
enough to overcome the increase of the elastical energy.
Here it deserves to mention that the same property of the excitation spectrum
is also present for another dimerized state with the pattern
corresponding to a wavevector $(\pi,0)$, i.e., $u_{i,j}=(-1)^{i}u$, 
see Ref. \cite{Katoh}.
Thus it is reasonable to predict that the similar conditional SP
transition will occur if this different dimerization pattern is considered.

We emphasize that the above results were obtained under the 2D JW 
transformation, where the interchain and intrachain exchanges were treated 
on the same level although the phase factor was 
approximately averaged. One may expect that the effect of interchain 
coupling, i.e., favoring the existence of the U state, could be adequately 
considered in this treatment. Actually, this point has been 
reflected in Fig. \ref{fig_PhDU}, prominently by the relationship between
$\eta_c$ and $h$ in the region $h\rightarrow 0$: $\eta_c \sim -1/\ln h$.
Even an infinitesimal interchain coupling may induce a rapid increase 
of $\eta_c$ away from zero, i.e., the U state is largely stabilized 
once the interchain coupling is switched on.

As mentioned before, a natural extension is to consider the case of the
Heisenberg model. In this case the U state should imply $z$-axis (and 
rotationally invariant) antiferromagnetic 
long-range order (LRO) \cite{Sandvik}. But we believe that the above 
qualitative results for the XY model, especially the rapid increasing 
property for $\eta_c$ away from zero,
is not changed by the extra Ising terms except that the U phase in 
Fig. \ref{fig_PhDU} is replaced by more precisely called N\'eel phase 
(i.e., uniform state with antiferromagnetic LRO) 
\cite{Fukuyama,Dobry}. More work is necessary for justification.

Before closing we note that recently there are intensive studies on the 
material CuGeO$_3$ which show surprising phenomena under doping \cite{Masuda},
and theoretically it has been understood that the interchain 
coupling plays a key role for these phenomena \cite{Saito,Fabrizio}. For 
their better understanding our phase diagram given here 
(for the pure system), which address the effect of the interchain coupling, 
will be helpful. Furthermore it is of great interest to 
extend this work to study the phase diagram for doped quasi-1D SP system. 
In addition, we also
notice that the nonadiabatic effect as refered before is important for really
modelling CuGeO$_3$ \cite{Caron,Bursill,Dobry,Uhrig}, so inclusion of dynamic 
phonons beyond this work is another interesting topic for future investigation.

In conclusion, based on the 2D JW transformation we have studied the ground 
state D/U phase diagram for an anisotropic 2D XY model in adiabatic limit. 
It is found that the spin-lattice coupling $\eta$ must exceed some critical 
value $\eta_c$ in order to reach the D phase for any finite $h$. The value 
$\eta_c$ has the dependence $-1/\ln h$ on $h$ in the region $h\rightarrow 0$ 
and the transition between U and D phase is of first-order for at 
least $h>10^{-3}$.

\bigskip
One of authors (Q. Yuan) would like to thank P. Thalmeier, T. Yamamoto, 
M. Azzouz for helpful discussions and the support of Visitor Program of 
MPI-PKS, Dresden. This work was also partly supported by the Chinese NSF.


\begin{references}
\bibitem{Remark} When $\eta >4$ it does not give nonzero solution for 
   $\delta$ any longer. But this actually means that the system is now 
   in the maximal dimerized state (i.e., $\delta^*=1$).
\bibitem{Supp} It is difficult to discern whether the transition is of 
   first- or weak second-order in the extremely small $h$ region.
\end{references}
\end{document}